 \documentstyle[preprint,aps]{revtex}

\begin{document}

\title{Quantum Electro and Chromodynamics treated by Thompson's heuristic
 approach}

\author{Cl\'audio Nassif * and P.R. Silva}

\address{*Departamento de Fisica - ICEB- UFOP, Morro do Cruzeiro,
35400-000- Ouro Preto-MG.} 

\address{ Departamento de F\'{\i}sica - ICEx - UFMG
Caixa Postal 702 - 30.123-970 - Belo Horizonte - MG - Brazil
cnassif@fisica.ufmg.br
prsilva@fisica.ufmg.br}
\par
\date{\today}
\maketitle
\begin{abstract}

In this work we apply Thompson's method (of the dimensions and scales) to study
some features of the Quantum Electro and Chromodynamics. This heuristic method
 can be considered as a simple and alternative way to the Renormalisation Group
 (R.G.) approach and when applied to QED-lagrangian is able to obtain in a first
 approximation both the running coupling constant behavior of $\alpha (\mu)$ and
 the mass $m(\mu)$.The calculations are evaluated just at $d_c=4$, where $d_c$
 is the upper critical dimension of the problem, so that we obtain the
 logarithmic behavior both for the coupling $\alpha$ and the excess of mass 
$\Delta m$ on the energy scale $\mu$. Although our results are well-known in the
 vast literature of field theories,it seems that one of the advantages of
 Thompson's method, beyond its simplicity is that it is able to extract directly
 from QED-lagrangian the physical (finite) behavior of $\alpha(\mu)$ and
 $m(\mu)$, bypassing hard problems of divergences which normally appear in the
 conventional renormalisation schemes applied to field theories like QED.
 Quantum Chromodynamics (QCD) is also treated by the present method in order to
 obtain the quark condensate value. Besides this, the method is also able to
 evaluate the  vacuum pressure at the boundary of the nucleon. This is done by 
assumming a step function behavior for the running coupling constant of the 
QCD, which fits nicely to some quantities related to the strong interaction 
evaluated through the MIT-bag model. 
\end{abstract}

\narrowtext

\section{Introduction}

\hspace{3em}

There are a considerable number of problems in Science where fluctuations are
 present in all length scales, varying from microscopic to macroscopic
 wavelengths.

As examples, we can mention the problems of fully developed turbulent fluid flow, critical phenomena and elementary particle physics. The problem of non-classical reaction rates (diffusion limited chemical reactions) turns out also to be in this category.

As was pointed out by Wilson\cite{1}: ``in quantum field theory, ``elementary" particles like electrons, photons, protons and neutrons turn out to have composite internal structure on all sizes scales down to zero. At least this is the prediction of quantum field theory".

The most largely employed strategy for dealing with problems involving many length scales is the ``Renormalization - Group (RG) approach".
The RG has been applied to treat the critical behavior of a system undergoing second order phase transition and has been shown to be a powerful method to obtain their critical indexes\cite{2}.

 In Quantum Electrodynamics (QED), Gell-Mann and Low\cite{3} obtained a RG
 equation for electron charge $e_\mu$, being $\mu$ the energy scale, so that in the limit as $\mu$ goes to zero we obtain the classical electron charge {\sl e}, and as $\mu$ goes to infinity we get the bare charge of the electron $e_B$.

The differential equation evaluated by Gell-Mann and Low obtains the ``experimental" charge $e_\mu$ of electron as a function of the energy, which corresponds to an interpolation between the classical and bare charges, namely:
 $e< e_\mu < e_B$.

In an alternative way to the RG approach, C. J. Thompson\cite{4} used a 
 heuristic method (of the dimensions) as a means to obtaining the correlation length critical index ($\nu$), which governs the critical behavior of a system in the neighborhood of its critical point. Starting from Landau-Ginzburg-Wilson hamiltonian or free energy, he got a closed form relation for $\nu(d)$ \cite{4}, where d is the spatial dimension. It is argued that the critical behavior of this $\Phi^4$-field theory is within the same class of universality as that of the Ising Model.

 One of the present authors\cite{5} applied Thompson's method to study 
diffusion limited chemical reaction ${\bf A+A \to 0}$ (inert product). The
 results obtained in that work\cite{5} agree with the exact results of Peliti\cite{6} who renormalized term by term given by the interaction diagramms
in the perturbation theory.

More recently, Nassif and Silva\cite{7} proposed an action to describe diffusion limited chemical reactions belonging to various classes of universality. This action was treated through Thompson's approach and could encompass the cases of reactions like ${\bf A+B \to 0}$ and  ${\bf A+A \to 0}$ within the same formalism. Just at the upper critical dimensions of  ${\bf A+B \to 0}$ ($d_c=4$) and  ${\bf A+A \to 0}$ ($d_c=2$) reactions, the present authors found universal logarithmic corrections to the mean field behavior.

Thompson's renormalisation group method has been applied to obtain the correlation length critical exponent of the Random Field Ising Model by Aharony, Imry and Ma\cite{8} and by one of the present authors\cite{9}. His method was also used to evaluate the correlation length critical exponent of the N-vector Model\cite{10}. Yang - Lee Edge Singularity Critical Exponents\cite{11} has been also studied by this method.

The aim of this work is to apply Thompson's method to study some features 
of the QED and also QCD. As we will see, the evaluated $QED_4$ coupling $\alpha$
 and the renormalized electron mass $m$ will exhibit logarithmic corrections
 on the energy scale. Here we would like to make some "alert" as a form of caution: It is well known that the proper way of dealing with the problems of QED is by using the techniques of the Quantum Field Theory (QFT),by first treating the leading Feynman diagramms in the context of the perturbation theory. On the other hand, perturbation theory is not appropriate to deal with
the infrared behaviour of the strong interaction described by the QCD. 

 QFT has also developed the apparatus for dealing with the divergences such as clever regularization schemes which leads to the renormalization of physical quantities  to be measured. However the subject of this work is by far of more modest achievements.We have been exploring the various possibilities of the Thompson's method of dimensions (see refs [5][7][9][10][11]\cite{12}\cite{13}\cite{14}
 \cite{15}). As we can see , for instance, by considering these various
 possibilities of the method, we were able to obtain the universal logarithmic
 behavior for the coupling parameters of various models at their respective
 upper critical dimensions [5,7,9-14]. So now we would like to know how this
 method behaves when applied to $QED_4$ in order to obtain the logarithmic
 behavior on scale of coupling $\alpha$ (for $d=4$). The use of the present
 method here could be better justified taking into account the simplicity and
 universality of its application to study the behavior of coupling parameters 
of other models.

In section II, we start by first considering a heuristic prescription which characterizes Thompson's method\cite{4}. There we also introduce the QED lagrangian.

Third section is dedicated to some further elaboration of Thompson's method when applied to $QED_4$.
 
In the fourth section we study the equation describing the dependence both of the coupling constant $\alpha$  and the mass $m$ on the energy scale ($\mu$).
Finally the next one will be dedicated to Quantum Chromodynamics (QCD) where 
we will evaluate the value of condesate of quarks given in its fundamental state
 and also the vacuum pressure at the boundary of the nucleon. Such results are
 motivated by the MIT-bag model\cite{16}\cite{17}\cite{18}\cite{19}
\cite{20}, developed to describe the strong interaction inside the hadronic
matter (nucleons). 

The last section is dedicated to the conclusions and prospects.
\section{$QED$ Lagrangian under the Thompson's method viewpoint}

 In this section, we start by writting the physical QED lagrangian, namely:
 
\begin{equation}
L = i\overline\Psi\gamma^{\mu}\partial_{\mu}\Psi -     m\overline\Psi\Psi-\frac{1}{4}F^{\mu\nu}F_{\mu\nu}+
ie\overline\Psi\gamma^{\mu}A_{\mu}\Psi
\end{equation}

with

\begin{equation}
F^{\mu\nu} = \partial^{\mu} A^{\nu} - \partial^{\nu} A^{\mu} \quad and~~~~ \overline{\Psi}=\Psi^\dagger\gamma^\circ
\end{equation}

In(1) $\Psi$ are fermion fields, $e$ and $m$ are respectively the electron's 
 charge and rest mass, $A_\mu$ is the four-vector electromagnetic potential and $\gamma^\mu$ are the Dirac's matrices.

 We assume that a heuristic approach used by Thompson[4] to study critical 
phenomena can be applied to lagrangian (1). It states that: 

``When we consider the integral of the Lagrangian (1) in a coherence volume $l^d$ in d-dimensions, the modulus of each integrated term of it is separately of the order of unity".

This method was firstly applied by Thompson[4] to the Landau-Ginzburg-Wilson free  energy or Hamiltonian, obtaining critical exponents within the same universality class of the Ising model.

But in fact, when we consider the integrals of each term in (1) as of the order of unity, we are really making a certain scaling dimensional analysis in each term of it. In doing this we have performed some scaling averages obtained separately from each integrated term of the lagrangian. In order to best 
justify the Thompson's prescription[4],we make the following reasoning: It is
well known that $QED_4$ has a trivial fixed point ($\alpha\approx 0$) at long-
wavelength regimes ($\lambda\sim l\rightarrow\infty$). Therefore if we want to 
describe the running coupling constant at low energy scales, it seems to be a 
good hypothesis to consider the behavior of the $QED_4$ lagrangian in the 
neighborhood of its trivial fixed point. This scaling invariance,  
which is also shared by a cooperative system at the critical point has been 
used by Thompson[4] in order to obtain the correlation length critical index
of the Ising-like systems,obtained close to its non-trivial fixed point.
 Indeed it is also well known that such a scaling invariance is in agreement
with the canonical dimensions obtained here for the fields 
$[\Psi^2]_l(\sim l^{-3})$,$[A_{\mu}]_l(\sim l^{-1})$,and the coupling constant
 ($[\alpha]_l\sim l^0\sim constant$) close to the fixed point
($\alpha\approx 0$). So we conclude that the scaling prescription of Thompson,
which comes from a dimensional analysis, works well in the neighborhood of a given fixed point. 

We borrow Thompson's idea to apply it to $QED_4$, but in order to better do this, we make the following considerations:

As we consider the terms which appear in (1) to be quadratic forms like $[\overline{\Psi}\Psi]$ and $[F^{\mu\nu}F_{\mu\nu}]$, we are going to take Thompson's prescription into a more refined form, so that we can put the modulus
 of each integrated term exactly equal to the unity since these terms are taking
in equal-footing. Besides this we will perform the integrations in the four
 -dimensional (4-D) space-time. 

 On the other hand, the idea of dimensional analysis is very common and when 
  applied to evaluate the dimension of $L$ in QED leads to 
 "$[L]=l^{-d}=\Lambda^d$", in d-dimensions, where $l$ is the length
 and $\Lambda$ is the momentum. By applying this prescription to each term of
 $L$, we obtain from the first one the dimension of the
field $\Psi^2$, that is simply $[\Psi^2]=\Lambda^{d-1}$, which gives
$[\Psi^2]=\Lambda^3=l^{-3}$ for $d=4$.

In a similar way, from the third term of $L$ we get $[A^2_\mu]=\Lambda^{d-2}$,
 being $[A^2_\mu]=\Lambda^2=l^{-2}$ in the special case $d=4$.

 Thompson's approach is based on a dimensional analysis in the energy scale
 (scaling reasoning) plus some additional heuristic considerations which lead to
 some mean values on the scale $l$ for the field $\Psi, A_\mu$, mass and charge
 (coupling $\alpha$).

Now applying Thompson's scaling assumption to the first term of (1) we have the following scaling integral: 

\begin{equation}
\left|\int_{l^4}i[\partial_\mu][\overline{\Psi}\Psi]d^4x\right| = 1. 
\end{equation}

We can observe that the dimension of \'~$\gamma^\mu\partial_\mu$\`~                 ($[\gamma^\mu\partial_\mu]_l$) which would appear in the integral is the same
 as $[\partial_\mu]_l=l^{-1}$. This is because we are thinking only about a
 dimensional analysis in (3) for \'~$\gamma^\mu\partial_\mu$\`~. So in this case
 we can naturally neglect the spinorial aspect of the field and just consider
 the \'~first derivative $\partial_\mu$\`~, which defines the fermions regarding to the scaling dimensional analysis: $[\partial_\mu]_l=l^{-1}$. 

On the other hand, when we are dealing with scalar fields, the second derivative picks up the bosonic behavior in a scaling dimensional analysis, i.e,
$[\partial^{\mu}\partial_{\mu}]_l=[\partial_{\mu}^2]_l=l^{-2}$\cite{21}.   

We would like to stress that, in his treatment of the critical phenomena, Thompson \cite{4} has considered integrals of the kind given by (3) in a more general case of d-dimensions. However, our interest here is restricted to the four-dimensional case, which is the most relevant if we take into account relativity theory, namely QED in (3+1) dimensions.

It is interesting to note that the integral (3) leads immediately to a kind of
 scaling dimensional analysis, where the dimensional value of certain quantity $[\overline{\Psi}\Psi]$ inside the integral is taken out of its integrand as a mean value in a coherent hyper-volume scale $l^4$. Then, from 
(3) we extract the following scaling behavior:\\
$\left<[\overline{\Psi}\Psi]\right>_l \equiv [\overline{\Psi}\Psi]_l=
[\Psi^2]_l\sim l^{-3}$, which also corresponds to a mean value of
$[\overline{\Psi}\Psi]$ on scale $l$, where we have considered a 4-D hyper-cubic
 volume $l^4$ for (3), being $[\partial_\mu]_l=l^{-1}$.  

In order to apply Thompson's prescription to the second term of (1), we would consider the scaling $\left|-\int_{l^4}[m\overline{\Psi}\Psi]_x d^4x\right|= 1$. However, a close examination reveals that this procedure does not work quite well. Due to the coupling between the $\Psi$ and $A$ fields, it is the mass increment that must be considered in the above relation. By putting $\Delta m=m(\mu)-m_0$, being $\mu$ the energy scale ($\mu = l^{-1}$), $\Delta m$ goes to zero as $\mu \to 0$ (or equivalently $l \to \infty$). After this consideration we can write: 
\begin{equation}
\left|-\int_{l^4}[\Delta m]_x[\overline{\Psi}\Psi]_x d^4x\right| = 1. 
\end{equation}

It is worth to emphasize that relation (4) was written by making the requirement that the quantities involved in it must satisfy a scaling relation. Relation (4) implies that:
\begin{equation}
\left<[\Delta m]\right>_l\left<[\overline{\Psi}\Psi]\right>_ll^4=1~,
\end{equation}
\\
or simply
\begin{equation}
[\Delta m]_l[\overline{\Psi}\Psi]_ll^4=1. 
\end{equation}

As we know, $[\overline{\Psi}\Psi]_l$ goes as $l^{-3}$, such that,from (6), we get that $[\Delta m]_l$ scales as $l^{-1}$, i.e, $[\Delta m]_l\sim l^{-1}$. 
 
To use Thompson's assumption in the third term of (1), it is better to think 
in terms of scaling behavior for the density of energy in the electromagnetic field ($\rho$), that is to say:

\begin{equation}
\frac{1}{8\pi}\int_{l^4}([E^2]_x +[B^2]_x)d^4x = 1~,
\end{equation}

Relation (7) implies that $[E^2]_l = [B^2]_l \sim l^{-4}$. We know that $\vec B=\nabla \times \vec A$. So making a dimensional analysis for $A$, we obtain: $[A^2]_l=[B^2]_ll^2 \sim l^{-4}l^2=l^{-2}$.
 
  Now let us consider the last term of the lagrangian (1). At a first sight, we
 could take a scaling integral for this term on a 4-D hyper-sphere($\sim l^4$).
Once the field $A_{\mu}$ in this term is a  fluctuating field due to the fact that photons are emitted and absorbed by the electron, the mean value of $A_{\mu}$ on scale of suficiently long times would vanishes ($\left<A_{\mu}\right>=0$). So we need to consider something quadratic like $\left<A^2_{\mu}\right>$ in order to avoid null mean value, or in other words, we need to evaluate a kind of second moment of this quantity. So based on this reasoning, as a means to extract a physical information (a quadratic coupling like $e^2=\alpha$) on the last term of (1), we judge necessary to perform a kind of second moment for the interaction term by considering an effective
contribution for the action through a product of integrals, i.e, we must look 
for a product of integrals in a 4-dimensional space, which corresponds to an 
average of the square of the last term of (1) in an effective space of 8-
dimensions. Making these considerations we firstly write

\begin{equation}
\left|i^2\int_{l^{\prime 4}}[\int_{l^4}(e[\overline\Psi\Psi]A_{\mu})d^4x]
(e^{\prime}[\overline\Psi^{\prime}\Psi^{\prime}]A^{\prime}_{\mu})d^4x^{\prime}
\right|=1,
\end{equation}
where '$\prime$` is a dummy index. 

 We can also write (8) in the following way: 

\begin{equation}
\left|i^2\int_{l^8}(e^2[\overline{\Psi}\Psi]_x^2[A^2_\mu]_x)d^8x \right| = 1~,
\end{equation}

which simply represents the following scaling: $[(e[\overline{\Psi}\Psi]_l[A_{\mu}]_l)l^4]^2=1$.  

\section{Some further elaboration of Thompson's method applied to $QED_4$.}

\hspace{3em}

  By reasons of spatial symmetry, let us now consider integral (3) evaluated
 in a volume of a 4 - D hyper-sphere, once we are interested in the isotropic 
4-D space-time, being the scale of length $l$ the radius of this hyper-sphere.

 The volume of a n-D hyper-sphere is given by $V_n = S_n\frac{l^n}{n}$\cite{22} where $S_n =\frac{2\pi^{\frac{n}{2}}}{\Gamma(\frac{n}{2})}$\cite{22}. For 4-D, we have $V_4 = \frac{\pi^2l^4}{2}$, implying  $dV_4 = 2\pi^2 r^3 dr$, where $r$
 is a radial variable.

The above considerations permit us to write integral (3) as being
  
\begin{equation}
   2\pi^2\int_{0_{(V_4)}}^l \partial_r [\overline{\Psi}\Psi]_r r^3dr =       \left<[\overline{\Psi}\Psi ]\right>_l 6\pi^2\int_{0}^l r^2dr=1. 
\end{equation}

Equation (10) implies that \\
  \begin{equation}
    \left<[\overline{\Psi}\Psi] \right>_l \equiv [\overline{\Psi}\Psi]_l = \frac{1}{2\pi^2l^3}~,
   \end{equation}
where $2\pi^2l^3$ is the magnitude of the surface of this 4-D hyper-sphere centered over the point charge $e$ whose "field" (fermionic field) is given by the amplitude $[\overline\Psi\Psi]_l$ above. 

 "$[\overline{\Psi}\Psi]_l$" could be thought of as a mean of the squared 
fermionic field where the average is taken on a length scale $l$, being $l=\mu^{-1}$. Therefore this squared fermionic field has the dimension of
 $\mu^3$ (the third power of energy). The "$2\pi^2$" constant is a simple
 consequence of the spherical symmetry we have assumed for the problem.

Now let us  evaluate  the mass term given by integral (4) in the volume of a 
4-D hyper-sphere of radius $l$. So we write

 \begin{equation}
 \left|-2\pi^2\int_{0}^l[\Delta m]_r[\overline{\Psi}\Psi]_r r^3dr\right| = 1~.
 \end{equation}

Relation (12) implies that

 \begin{equation}
 \left<[\Delta m]\right>_l \left<[\overline{\Psi}\Psi]\right>_l \frac{\pi^2 l^4}{2} = 1~.
 \end{equation}

By putting (11) into (13), we get:

 \begin{equation}
 \left<[\Delta m]\right>_l \equiv [\Delta m]_l = 4l^{-1}. 
 \end{equation}

Now let us consider the third term of lagrangian (1) and by performing the integral (7) in the volume of a 4-D hyper-sphere of radius $l$ ($dV_4=2\pi^2r^3dr$),where the squared fields have the same scaling behavior $[E^2]_l=[B^2]_l\sim l^{-4}$, so we write:

 \begin{equation}
  \frac{\pi}{2}\int_{0}^l[E^2]_r r^3dr = \frac{\pi}{2}\int_{0}^l[B^2]_r        r^3dr =1. 
 \end{equation}

Relation (15) leads to

  \begin{equation}
  [E^2]_l  = [B^2]_l = \frac{8}{\pi l^4}(\sim l^{-4}). 
 \end{equation}

From the definition  $\vec B=\vec \nabla \times \vec A$, we are led to the following scaling relation for $[A^2]_l$:

 \begin{equation}
  [B^2]_l l^2  = [A^2]_l = \frac{8}{\pi l^2}~~,  
 \end{equation}
where in obtaining (17), we have used (16).

It is interesting to note that (17) is consistent with a potential of a static point charge, that is to say $\Phi \sim \frac{1}{r}$, which leads to $[\Phi^2]_l \sim \frac{1}{l^2}$, where $\Phi \equiv A_4$ and $A_{\mu} = (\vec A, \Phi)$. These considerations permit us to write (17) in a compact form:
 
 \begin{equation}
  [A^2_{\mu}]_l   =  \frac{8}{\pi l^2}~(\sim l^{-2}).
 \end{equation}

With respect to the fourth term of $L$ in (1), previous considerations had led
 to integral given by (9). Therefore let us now evaluate integral (9) in
 a volume of a 8-D hyper-sphere. Taking into account that $d^8x \equiv dV_8 =
 S_8r^7dr=\frac{\pi^4}{3}r^7dr$, we have:
  
 \begin{equation}
 \frac{\pi^4}{3}\int_l[\alpha]_r[\overline{\Psi}\Psi]^2_r [A^2_{\mu}]_rr^7dr = 1~,
 \end{equation}
\noindent
where $\alpha = e^2$.

A first trying in order to evaluate (19) could be to write it as a product of averages, namely as a product of the quantities $\left<[\alpha]\right>_l, \left<[\overline{\Psi}\Psi]^2\right>_l,  \left<[A^2_{\mu}]\right>_l$ and $[V_8]_l$. By considering that $\left<[\overline{\Psi}\Psi]^2\right>_l \sim l^{-6}$, $\left<[A^2_{\mu}]\right>_l \sim l^{-2}$ and $[V_8]_l \sim l^8$, we obtain that $[\alpha]_l l^{-6} l^{-2} l^8 \sim 1$, which implies that $[\alpha]_l$ is a constant, that is to say a quantity which does not exhibit a dependence on the scale of length $l$ (or energy $\mu$): $[\alpha]_l \sim l^0 \sim$ constant (scale invariance over the trivial fixed point).

 We must consider that '$d=4$` corresponds to a kind of upper critical dimension for QED. In other words, below $d=4$ , fluctuations are very important to the problem, and above $d=4$, ``mean field" description \cite{7} is a good description to the problem. So $d=4$ represents a border-line dimension for QED and we must improve our approximations in order to ``see" the logarithmic dependence of the coupling $[\alpha]_l$ on the length scale $l$, or equivalently on the energy scale $\mu=l^{-1}$. It must be stressed that a similar situation has been occurred when we treated diffusion limited chemical reactions through Thompson's method \cite{5,7,12,13,14}.
As a means to improve the calculation of (19) let us take the quantities $[\overline{\Psi}\Psi]^2_r$ and $[A^2_{\mu}]_r$ inside the integral with the same form as those evaluated in (11) and (18), but now displaying a dependence
 on the r-variable of scale. So by taking inside the integral (19)
 $[\overline{\Psi}\Psi]_r = \frac{1}{2\pi^2 r^3}$ and $[A^2_{\mu}]_r = \frac{8}{\pi r^2}$, we can write: 
  
\begin{equation}
 \frac{\pi^4}{3}\int_{l}[\alpha]_r \left(\frac{1}{2\pi^2r^3}\right)^2 \left(   \frac{8}{\pi r^2}\right) r^7dr = 1. 
\end{equation}

From (20),by putting the mean value of $\alpha$ on scale $l$ out of this
 integral, we have:

\begin{equation}
 \left<[\alpha]\right>_l\frac{2}{3\pi}\int_{1}^l \frac{dr}{r} = \left<[\alpha]\right>_l \frac{2}{3\pi}ln(l) = 1.  
\end{equation}

In evaluating (21), we have taken  1 as a lower cutoff on the scale $l$. Therefore (21) displays the logarithmic dependence for $[\alpha]_l$ on the scale
 of length $l$ (or energy $\mu = l^{-1}$).

\section{Evaluation of the dependence of charge and mass of the electron with the scale of energy}

 \subsection{Obtaining $\alpha(\mu)$}
\hspace{3em}

 In the quantum regime (vacuum polarization), the behavior of $\alpha$ is given by Eq.(21).\\ 
For the sake of simplicity in the notation, we write:

\begin{equation}
 \left<[\alpha]\right>_l \equiv [\alpha]_l \equiv \alpha(l)~~,
\end{equation}

and by putting $\mu=l^{-1}$ into (21), we get 

\begin{equation}
 -\frac{2}{3\pi}ln(\mu)=\alpha^{-1}(\mu)~.
\end{equation}

Differentiating both sides of (23) with respect the $\mu$ variable, we obtain:

\begin{equation}
 \mu \frac{d\alpha}{d\mu}=\frac{2}{3\pi}\alpha^2~.
\end{equation}

 Equation (24) coincides with that which is obtained by the R.G procedure, when  $QED_4$ is treated through the perturbation theory at one loop level.We remember
 that, in performing the calculations by using the first prescription of
 Thompson's method, we firstly derived a coupling constant which does not depend
 on the energy scale. As a means to seek for a running coupling constant, we
 need to make a fine-tunning (see equation (20)) which recovers some
 fluctuations corrections which are present on scales of moderate energy. For
the case of higher energy scales, we have already made an estimative for running
 coupling constant\cite{23} taking into account some additional assumptions for
Thompson`s prescription in order to get stronger fluctuations corrections. 
 
  The idea of fine-tunning was also used before in some previous works
 [5],[7],[12-14], and it allowed us to obtain simple logarithmic corrections on
 energy scale just at the upper critical dimensions of those models[5,7,12-14].
 
  From (24) we observe that we have obtained the coefficient
 $\beta=\frac{2}{3\pi}\alpha^2$ \cite{24},\cite{25}.

Performing the integration of (24), by considering the limits $\mu_0$ and $\mu$ for the energy scales and their respectives couplings $\alpha(\mu_0)$ and $\alpha(\mu)$, we obtain:

\begin{equation}
\alpha(\mu) =\frac{\alpha(\mu_0)} {1-\frac{2}{3\pi}\alpha(\mu_0)ln\left(\frac{\mu}{\mu_0}\right)}. 
\end{equation}

We observe that (25) diplays the so-called Landau's singularity, namely a finite value of the energy scale $\mu_L$, where $\alpha(\mu_L) \to \infty$.

As it is well known, Landau's singularity is a non-physical effect and reveals the fact that the running coupling constant solution given by (25) is not appropriate when the energy scale approaches $\mu_L$. 

We are led to think that at higher energies, equation (24) and its solution (25) must be modified in order to be free of the  Landau's singularity. In the usual perturbative scheme of calculation this is accomplished by considering the theory beyond one loop level (two or more loops).

Now let us look at (25). We observe that $\lim_{\mu \to 0} \alpha (\mu)=0$. But this result seems to be purely of academic interest.

Indeed even at low energy scales, the departure of the classical behavior for $\alpha(\mu)$ starts when $\mu > m_0$, where $m_0$ is the electron rest mass. This corresponds to assume that the effect of vacuum polarization in the shielding of electron charge becomes important when we approach  the electron closest than its Compton wavelength $l_0 \sim \lambda_c = m^{-1}_0$. Therefore, from an experimental point of view, we must look for (25) at the lower energy scale regime, but with $\mu\geq\mu_0 \sim m_0$, that is, we consider the parameter of energy scale $\mu$ fixed on the electron mass $m_0$ as a scale of reference, where $\alpha(\mu_0)\sim \alpha(m_0)\approx \frac{1}{137}$. So for moderates energies we can make the expansion of (25), obtaining 

\begin{equation}
 \alpha(\mu) = \alpha_0\left[1 + \frac{2}{3\pi}\alpha_0 ln\left(\frac{\mu}{\mu_0}\right) \right]~~,
\end{equation}
\noindent 
where  $\alpha_0 = \alpha(\mu_0) \sim \alpha(m_0)\approx \frac{1}{137}$, and $\mu_0 \sim m_0$.

The result above (26) is well known in the literature. See reference\cite{26}. 

\subsection{Attainment of $m(\mu)$}
\hspace{3em}
As a means to evaluate $\Delta m(\mu)$ let us  compare (4) and (8) by considering the shift $\Delta e^2=\Delta \alpha$ in (8) because $\Delta \alpha (\mu)$ must be directly proportional to the mass shift given in (4), that is, $\Delta m \propto \Delta \alpha$ so that in the very lower energies limit we have $\Delta m \propto \Delta \alpha = \Delta e^2 \to 0$. Thus in doing that we obtain 

\begin{equation}
 \left|\int_{V_4}(-\Delta m)[\overline\Psi\Psi]_x d^4x\right|=   \left|i^2\int_{V_4}(\int_{V^{\prime}_4}(\Delta e^{\prime 2})[\overline\Psi
^{\prime}\Psi^{\prime}]_x^{\prime}[A_{\mu}^2]_x^{\prime}d^4x^{\prime})
[\overline\Psi\Psi]_x d^4x\right| = 1~~,
\end{equation}
where we consider the shift $\Delta m= m-m_0$ and
 $\Delta\alpha=\alpha-\alpha_0$, being $m_0$ the electron rest mass and 
$\alpha_0\approx\frac{1}{137}$ a constant measured on energy scale of electron 
rest mass. 

 Relation (27) implies that 

\begin{equation} 
\Delta m=\left|i^2\int_{V_4}(\Delta e^2)[\overline\Psi\Psi]_x[A_{\mu}^2]_xd^4x
\right|,
\end{equation}
where the index '$\prime$` is dummy. 

 By putting $d^4x\equiv dV_4=2\pi^2r^3dr$, $[\overline\Psi\Psi]_x\equiv
[\overline\Psi\Psi]_r=\frac{1}{2\pi^2r^3}$ and $[A_{\mu}^2]_x\equiv
[A_{\mu}^2]_r=\frac{8}{\pi r^2}$ (see (11) and (18)) into (28), we get 

\begin{equation}
\Delta m= \frac{8}{\pi}\left<[\Delta\alpha]\right>_l\left|\int_l\frac{1}{r^2}dr
\right|. 
\end{equation}

 Now let us take the notation $\left<[\Delta\alpha]\right>_l\equiv\Delta\alpha$
to represent a mean charge shift measured on the scale $l\sim\mu^{-1}$. Thus 
performing the integration (29) between the limits $\mu=0$ ($l=\infty$) and
$\alpha_0 l_0\sim\alpha_0\lambda_c\approx 10^{-14}m$, which is equivalent to the
vacuum polarization regime due to the small value of length scale $\alpha_0\l_0$
($<<\lambda_c$,i.e, the Compton wavelength), we obtain 

\begin{equation}
\Delta m=\frac{8}{\pi}\Delta\alpha\frac{1}{\alpha_0 l_0}=\frac{8}{\pi}
\frac{\Delta\alpha}{\alpha_0}m_0
\end{equation}
where $m_0=l_0^{-1}\sim\lambda_c^{-1}$. Indeed we verify such a proportionality
"$\Delta m\propto\Delta\alpha$" (30) mentioned before, which leads to 

\begin{equation}
m=m_0 + \frac{8}{\pi}\frac{\Delta\alpha}{\alpha_0}m_0, 
\end{equation} 
where $\Delta m=m- m_o$.

 Finally by substituting  $\Delta \alpha(\mu)$ obtained from (26) into (31)
 we get

\begin{equation}
 m= m_0 \left[1+\frac{16}{3\pi^2}\alpha_0 ln\left( \frac{\mu}{\mu_0}\right)     \right]. 
\end{equation}
\noindent 

 We notice that the above relation is comparable to the result for $m(\mu)$ of the literature as quoted by Nottale\cite{27}, Weinberg\cite{28} and  Weisskopf
 \cite{29}. So, in spite of all results we have just obtained are well-known in
 the literature, we must stress again that the novelty here consists in
 obtainning another and a new way to deal with some problems of QED by applying
 Thompson's method of scales. 

It is also interesting to verify that the quantum correction to the electron
 mass could be obtained in a way which is consistent with (32) through the following reasoning: Let us consider the energy stored in the eletric field, namely  $U_ {el}=\int_{V_{3}} E^2 dV_{3}$ , where the integration is performed in a 3-D volume. Now let us write the eletric field $E$ as its classical value $E_0$ plus a correction $\Delta{E}$ due to the quantum fluctuations. We assume that quantum fluctuations only affects the energy $U_{el}$ through the squared contribution in $\Delta{E}$, once the linear term in $\Delta{E}$ averages out to zero. So we have $\overline{E^2} = \overline{E_0^2} +\overline{\Delta E^2}$ , where the bars means averaging over a suficiently long time in the scale of the fluctuations. Therefore, as we are mainly interested in the quantum process namely the absortion and emission of virtual photons, we can write 
  \begin{equation}
 \Delta{E_{rms}} = {[\overline{(\Delta E^2)}]}^\frac{1}{2}, 
\end{equation}
\noindent 

where the index $rms$ means root mean square. 

It is natural to think that $\Delta{E_{rms}}$ will be different from zero only in the presence of the fermionic field, and this leads us to propose the
 relation
  \begin{equation}
\Delta{E^2_{rms}}=\xi^2\Psi^2_{rms},
\end{equation}
\noindent

where we have considered $\Psi^2_{rms} = \left<[\overline\Psi\Psi]\right>_r=\frac{1}{2\pi^2r^3}$(see (11)),that is to say
,$\Psi^2_{rms}$ corresponds to the mean squared fermionic field in the variable of scale-$r$, and $\xi$ is a constant.
 
 It seems that the intuitive reasoning given in (34) is consistent with gauge 
invariance of the theory. That is, in order to take into account the  quantum fluctuations contribution to the electric field, we must have necessarily the coupling between the electric and the fermionic fields.  

Equations (34) and (11) imply that 

\begin{equation}
\Delta{E_{rms}} \propto \frac{1}{r^{\frac{3}{2}}},
\end{equation}
\noindent

which must be compared with the inverse square Gauss law of the classical contribution. At this point we would like to notice that a dependence of the quantum fluctuations of the eletric field on the scale of length as that we obtained in (35) was proposed by Weisskopf[28]. 

Now performing the integration of $\Delta E^2_{rms}$ in a 3-D volume and taking as integration limits the variable r $(r\leq\lambda_c)$ and $\lambda_c$ the Compton wavelength, we obtain 

\begin{equation}
(\Delta{m})c^2 = mc^2 - m_0c^2 = \int_r^{\lambda_c}(\frac{\xi^2}{2\pi^2r^3})4  {\pi}r^2 dr.
\end{equation}

The reason to consider $\lambda_c$ as long wavelength cutoff is that quantum fluctuations does not contribute very much to the electromagnetic mass of the electron above this value. 

Relation (36) implies that 

\begin{equation}
mc^2 = m_0c^2 + Cln(\frac{\lambda_c}{r}),
\end{equation}
\noindent

where $C$ is a constant. 

It is worth to emphasize that (37) is consistent with the results we have
 obtained in (32) and that by Weisskopf[28],if we fix $C\equiv\frac{3}{2\pi}
 m_0c^2\alpha_0$, being $\frac{\lambda_c}{r}\sim\frac{\mu}{\mu_0}$. 
\section{Thompson`s method, QCD and MIT-bag model} 

 Quantum Chromodynamics (QCD), the modern theory of the strong interactions
\cite{30}\cite{31} is a non-Abelian Field Theory. In 1973, Gross and Wilczek
\cite{32} and independently Politzer\cite{33} have shown that certain classes
of non-Abelian fields theories exhibit asymptotic freedom, a necessary condition
for a theory which could describe strong interactions. These seminal papers
[32,33] open the route to the birth of the QCD. 

 In a not very accurated picture, QCD can be considered as an expanded version  
of QED. In QCD we have also six fermionic fields representing the various 
quark flavors, in contraposition to a simgle fermionic field of the QED. Besides
the asymptotic freedom exhibit at the ultraviolet limit, a theory of the strong
interactions must also display quark confinement at the infrared limit.  

 Whereas in QED there is just one kind of charge, QCD has three kinds of charge, labeled by"color" (red, green and blue)[30]. The color charges are conserved in all physical process. There are also photon-like massless particles, called color gluons, that respond in appropriate ways to the presence of color charge. This mechanism is very similar to the ways photons respond to electric charge in QED.

 In QCD, quarks are particles that carry color charge. As we already know,
there are six different kinds of quarks, called "flavors", denoted by $u$ (up),
 $d$ (down); $c$ (charmed), $s$ (strange); $b$ (botton) and $t$ (top). Of these,
 only $u$ and $d$ quarks play a significant role in the structure of ordinary
 matter. They carry fractional electric charge, i.e, $+\frac{2}{3}e$ for $u$,
 $c$ and $t$ quarks, and $-\frac{1}{3}e$ for $d$, $s$ and $b$ quarks, in
 addition to their color charge. 

 In a similar way to QED-Lagrangian, let us write the QCD-Lagrangian density,
namely: 

  \begin{equation}
L=\Sigma_j \overline q_j(i\gamma_{\mu}D^{\mu}- m_j)q_j - \frac{1}{4}G^a_{\mu\nu}
G_a^{\mu\nu},
  \end{equation}

where $D^{\mu}=\partial^{\mu}+\frac{1}{2}ig\lambda_a A^{\mu}_a$ , and 
$G^{\mu\nu}_a=\partial^{\mu}A^{\nu}_a-\partial^{\nu}A^{\mu}_a-gf_{abc}A^{\mu}_b
A^{\nu}_c$\cite{34} 

 In (38) above, $m_j$ and $q_j$ are the mass and quantum field of the quark of 
$j^{th}$ flavor, and $A$ is the gluon field, being $\mu$ and $\nu$ the space- 
time indices. $a$, $b$ and $c$ are color indices. The numerical coefficients 
$f$ (structure constants) and $\lambda_a$ guarantee $SU(3)$ color symmetry. $g$
is the coupling constant. 

 By applying Thompson's assumption to the Knetics term from (38), we obtain a 
similar result as given before in QED (see equation (11)), namely 

 \begin{equation}
  \left<[\overline q_j q_j]\right>_l\equiv [\overline q_j q_j]_l=
-\frac{1}{2\pi^2l^3}.
  \end{equation}

 But now, the amplitude above must be interpreted as a quark condensate scaling,
and we took the negative signal in order to be consistent with the description 
of a bound state. 

  On the other hand, let us remember that we have obtained the mass shift 
$\Delta m$ of the electron in QED (see equation (29)). This is due to the 
interaction with electromagnetic field (photon). Following this same reasoning
for QCD, where we have quarks (fermions) interacting with gluon fields (bosons),
thus according to MIT-bag model idea, now let us think of a special behavior for
the running coupling constant ($\alpha_{running}$) which obeys a step-like
function specified by considering the boundary condition at the surface of
the bag, namely: 

 \begin{equation}
\alpha_{runing}=\left\{
\begin{array}{ll}
1, &  \frac{1}{m_0}<r<\infty\\ 
0, & r<\frac{1}{m_0}(= r_0), 
\end{array}
\right.
\end{equation} 
where $m_0$ is a reference mass above which the running coupling constant 
vanishes, i.e, for a radius below nucleon radius ($r<r_0$) or $m>m_0$,the quarks
have free motion inside the bag (the nucleons); so for $r>r_0$ we consider 
$\alpha=1$. 

 Finally, taking into account $\alpha_{running}$ in (40) inside the integral 
for $\Delta m$ (equation (29)), and making the integration between the limits
$r=\infty$ and $r=r_0=\frac{1}{m_0}$, we write
 
\begin{equation}
\Delta m =  
 \frac{8}{\pi}\left|\int_{\infty}^{\frac{1}{m_0}}\alpha_{running}\frac{1}{r^2}
dr\right|, 
\end{equation}

where we obtain

\begin{equation} 
\Delta m=\frac{8}{\pi} m_0.
\end{equation}

 For the special case of strong interaction ($\alpha=1$), we consider that
 the shift of mass, which comes from the strong interaction effects inside the
 nucleon, is pratically responsable for almost all the mass of nucleon. So due 
to this fact, we can make the following approximation: $\Delta m\equiv
 m_{nucleon}=m_n$. Thus by introducing it inside (42), we get    

\begin{equation} 
m_n=\frac{8}{\pi}m_0.  
\end{equation}

 Now using $h=c=1$, and also by considering $m_0$ as a zero-point energy inside
(43), i.e, we have $m_0=\frac{1}{2}\nu_0=\frac{1}{2l}$ inside (43), thus we 
obtain:
\begin{equation} 
m_n= 3m_q= \frac{4}{\pi l},
\end{equation}
where $m_q$ represents the constituent mass of quark ($m_q=\frac{1}{3}m_n$). 

 From (44) above, we get $\frac{1}{l}=\frac{3\pi}{4}m_q$. So by putting this 
result inside the scaling relation for the quark condensed given in (39), we 
obtain 
   \begin{equation}
  \left<[\overline q_j q_j]\right>_l\equiv [\overline q_j q_j]_l=-\frac{27\pi}
{128}m_q^3.
  \end{equation}

 Having $m_n\approx 939 MeV$, which implies that $m_q=\frac{1}{3}m_n\approx 
313MeV$, we finally get $[\overline q q]\approx -(\frac{m_q}{1.147}
 MeV)^3\approx -(273 MeV)^3$. 

 The value of the quark condensate evaluated above must be compared with other
theoretical values which go from $(-265 MeV)^3$ to $(-340 MeV)^3$ (see table (2)
in the paper by Mota $et al.$\cite{35}) and also with recent experimental 
result\cite{36} of $[-(296\pm 25)MeV]^3$.  

 According to MIT bag-model[17], the simplest shape for a bag is naturally a 
sphere (spherical bags), i.e, $R(\theta ,\phi)=R=constant$; ${\bf n}(\theta
 ,\phi)={\bf e}_r$ is the unitary normal vector to the spherical surface.
$R$ is the bag radius ($R\equiv r_0$). Another boundary condition is obtained 
by demanding that the pressure of the quarks on the bag surface be constant and
must equal a constant exterior vacuum pressure, i.e, we have $B=\frac{1}{2}
{\bf n}.{\bf\nabla}\Sigma_q \overline q q]_{R=R(\theta ,\phi)}$[17].  
 
 On the other hand, as we already know, the condensed $[\overline q q]$ was 
obtained by Thompson's scaling reasoning (equation 39). So inserting (39) into 
the formula above[17] from MIT-bag model, and by considering ${\bf n}. 
{\bf\nabla}= \frac{\partial}{\partial r}$, thus we get 

\begin{equation}
B= \frac{1}{2}\frac{\partial}{\partial r}[\overline q q]_r =
\frac{1}{2}\frac{d}{dr}(-\frac{1}{2\pi^2 r^3})=
\frac{3}{4\pi^2 r^4}. 
\end{equation}

 Having $hc=2\pi\hbar c=1$, then by introducing this information into (46)
above, we finally write

 \begin{equation}
B=\frac{6\hbar c}{4\pi r^4}. 
 \end{equation}

 The above result, which comes from MIT-bag model[17] in addition to the scaling
result for quark condensate in spirit of Thompson's approach (equation (39)), 
coincides with that obtained in another work, dealing with the MIT-bag
 model[16]. 

 Since nucleon has a radius in order of $10^{-15}m$, thus from (47) we obtain 
$B\sim 10^{29} atm$, which also represents an external vacuum pressure over the
nucleon. Such a pressure must equal the pressure inside the bag (nucleon) due   
to the quarks.

\section{Conclusions and prospects}
\hspace{3em}
In this paper, Thompson's heuristic method which could be considered as a simple
 alternative way to the RG calculations  was applied to study $QED_4$.This was
 done by treating each term of the QED lagrangian in equal footing, through  
 dimensional analysis on the scale of length (or equivalently on the
 momentum-energy scale). If we analyse the scaling behavior of certain objects
 such that the mean squared fermionic field ($[\overline{\Psi}\Psi]_l$), the
 dimension of the squared vector potencial $[A^2_\mu]_l$, the ``excess" of mass
 $[\Delta m]_l$, and the ``excess" of charge $[\Delta \alpha]_l$,  with all
 these quantities evaluated at the scale of length $l$, we observe that it is
 possible to organize these objects within a hierarchical structure, thinking in
 terms of topological grounds. In this way, the mean squared fermionic field
 $\left< [\overline{\Psi}\Psi] \right>_l = (2\pi^2 l^3)^{-1}$ decreases as a
 ``surface" 3-D of a hyper sphere 4-D of radius $l$, being this ``surface"
 immersed in the 4-D space-time.

The next object in this hierarchy corresponds to the dimension of the squared vector potential. It is given by $[A^2_\mu]_l = 8(\pi l^2)^{-1}$, exhibiting a inverse square law on the scale of length $l$. This represents a 2-D structure also immersed in the 4-D space-time. We could think that, for this object, the degree of freedom have reduced by a unity. The ``excess" of mass $[\Delta m]_l=4l^{-1}$ can be thougth of as a 1-D structure immersed again in a 4-D space-time.

Finally the ``excess" of charge (coupling) $[\Delta \alpha]_l$ behaves in the
 zero-th order as scaling independent, namely $\alpha$ goes as $l^0$ at zero order in the calculations, and it can be considered as a 0-D structure immersed in a 4-D space-time. In short, we have the ``spreading" of the squared fermionic
 field in a 3D-volume , the squared vector potential in a 2D-surface , the mass
 in a 1D-line  and the charge in a point (0-D), relating these objects of QED to
 a hierarchical ordering in the topology of a 4-D space-time.

However when we  improve our calculations, the charge (running coupling `constant') passes to exhibit a logarithmic dependence on the scale of length. This could be considered as an intermediate regime between a point ($l^0 \sim$ constant) and a line ($l^1$). We interpret this as the charge accquiring a fractal character in this topological structure of the space-time, due to the influence of the quantum fluctuations introduced by  the vacuum polarization, in such a way that we have $\alpha(l) \sim [ln(l)]^{-1}=[l^0 ln(l)]^{-1}$.
These quantum fluctuations will also ``modulate" the behavior of the ``excess" of mass, namely $\Delta m(l) \sim [l^1 ln(l)]^{-1}$.

The fractal character of a quantum path was considered by Nottale\cite{37} on analysing the QED. He showed that due to the vacuum polarization the self-energy diagramms of the QED display a fractal character[37]. 

One merit of Thompson's approach is that it displays the scaling behavior of the physical magnitudes of the problem, and as a consequence, it allows us for instance to pick up the fractal structure of $\alpha (\mu)$ and $m(\mu)$
(logarithmic dependence on scale).

 One of the possibilities of the Thompson's method is to use it as a means to
 evaluate the running coupling constant of Quantum Chromodynamics (QCD). This
 matter will be treated elsewhere.  
 
Finally, since Thompson's method is essentially a scaling approach, we can also apply it to study the growth of polymer chains in an alternative way to the R.G scheme. This matter will be also treated elsewhere.

\end{document}